# Can a mass inversion save solar neutrino oscillations from the Los Alamos neutrino?

Georg Raffelt
Max-Planck-Institut für Physik
Föhringer Ring 6, 80805 München, Germany

and

Joseph Silk
Astronomy and Physics Departments,
and Center for Particle Astrophysics
University of California, Berkeley, CA 94720, U.S.A.

15 February 1995

**Abstract**

In the light of the $\nu_\mu \to \nu_e$ neutrino oscillations which may have been observed at the LSND experiment we explore the consequences of two inverted mass schemes where solar neutrino oscillations occur between $\nu_e$ and $\nu_\tau$. The favored LSND value $\Delta m^2 = 6\,\text{eV}^2$ leads to $m_{\nu_e} \approx m_{\nu_\tau} \approx 2.5\,\text{eV}$ and $m_{\nu_\mu} \approx 0$ so that cosmology can benefit from a recently proposed "cold plus hot dark matter" structure formation scenario with two equal mass light neutrinos ($C\nu^2 DM$). Solar neutrino oscillations ($\nu_e \to \nu_\tau$) can occur with one of the large mixing angle solutions so that a serious conflict with $\beta\beta$ decay Majorana mass limits is avoided without invoking Dirac masses. However, there is a problem with the SN 1987A signal because of resonant $\overline{\nu}_e \leftrightarrow \overline{\nu}_\mu$ oscillations which are expected to cause far higher $\overline{\nu}_e$ energies at the IMB and Kamiokande II detectors than have been observed. A small value $\Delta m^2 = 0.5\,\text{eV}^2$ at LSND, which allows for a relatively large $\nu_e$-$\nu_\mu$ mixing angle without conflicting with the KARMEN and BNL-E776 experiments, would indicate $m_{\nu_e} \approx m_{\nu_\tau} \approx 1.62\,\text{eV}$ and $m_{\nu_\mu} \approx 1.77\,\text{eV}$. This scheme of $C\nu^3 DM$ maintains, and even may improve, the essential cosmological model implications for large-scale structure, leaving no conflict with SN r-process nucleosynthesis. It may improve the discordance between the SN 1987A neutrino spectra inferred from Kamiokande II and IMB.

The LSND collaboration (Los Alamos) has recently confirmed early rumors that they were seeing evidence for $\nu_\mu \to \nu_e$ oscillations [1], probably corresponding to $\Delta m^2 \approx 6\,\text{eV}^2$ and $\sin^2 2\theta \approx 0.005$ [2, 3]. This interpretation may be marginally consistent with previous exclusion areas of the BNL experiments 734 of Ahrens et al. [4] and 776 of Borodovsky et al. [5], as well as with the KARMEN experiment [6]. The atmospheric neutrino anomaly may be explained by $\nu_\mu \to \nu_\tau$ oscillations with $\Delta m^2 \approx 10^{-2}\text{eV}^2$ and nearly maximal mixing [7]. Taken together, these hypotheses hint at a neutrino mass matrix with $m_{\nu_e} \approx 0$ and $m_{\nu_\mu} \approx m_{\nu_\tau} \approx 2.5\,\text{eV}$. Primack et al. have shown that the presence of two 2.5 eV neutrinos as a "hot component" of the cosmic dark matter ("C$\nu^2$DM") may provide an elegant resolution of the structure formation crisis that bedevils a pure cold dark matter cosmology [3].

This explanation for results from two neutrino experiments seems attractive, but at least two significant, if not necessarily fatal, objections can be raised. Whether the atmospheric neutrino anomaly can indeed be explained by oscillations remains controversial as the best-fit parameters of Kamiokande [7] seem inconsistent with the exclusion plots of the Fréjus [8] and IMB collaborations [9]. Moreover, there is no room for solar neutrino oscillations $\nu_e \to \nu_\mu$ or $\nu_e \to \nu_\tau$ which need $\Delta m^2 \approx 10^{-5}\,\text{eV}^2$ (MSW solution) or $\Delta m^2 \approx 10^{-10}\,\text{eV}^2$ (vacuum solution). Thus, one needs to invoke a new sterile neutrino in order to explain the solar neutrino problem by oscillations, an ad-hoc assumption that we find very unsatisfactory.

Instead, one could give up on $\nu_\mu \to \nu_\tau$ oscillations as an explanation for the atmospheric anomaly and invoke $\nu_e \to \nu_\tau$ oscillations for the Sun. This implies $m_{\nu_e} \approx m_{\nu_\tau}$ and so, it would destroy the "cold plus hot dark matter" cosmology if $m_{\nu_e} \approx m_{\nu_\tau} \approx 0$ and $m_{\nu_\mu} \approx 2.5\,\text{eV}$. Also, it implies a mass inversion with $m_{\nu_\mu} > m_{\nu_\tau}$. Of course, if one is willing to contemplate a mass inversion one may equally consider the reverse situation with $m_{\nu_e} \approx m_{\nu_\tau} \approx 2.5\,\text{eV}$ and $m_{\nu_\mu} \approx 0$. For cosmology, this scenario would be equivalent to that of Primack et al. [3] while the solar neutrino problem is solved by $\nu_e \to \nu_\tau$ oscillations.

An additional benefit of the mass inversion between $\nu_e$ and $\nu_\mu$ is that it avoids a conflict [10] with the recent interpretation that r-process nucleosynthesis occurs in the neutrino-driven wind emanating from the surface of a nascent neutron star a few seconds after the core collapse in a type II supernova [11]. The neutron star emits neutrinos of all flavors with energies for which various authors find values in the ranges [12]

$$\langle E_\nu \rangle = \begin{cases} 10 - 12 \, \text{MeV} & \text{for } \nu_e, \\ 14 - 17 \, \text{MeV} & \text{for } \overline{\nu}_e, \\ 24 - 27 \, \text{MeV} & \text{for } \nu_{\mu,\tau} \text{ and } \overline{\nu}_{\mu,\tau}. \end{cases} \quad (1)$$

This hierarchy of energies is caused by the different opacities for the different species: they are emitted from different layers of the neutron star. The relation $\langle E_{\nu_e} \rangle < \langle E_{\overline{\nu}_e} \rangle$ implies that $\beta$ processes shift the wind to a neutron-rich phase necessary for the r-process. The LSND mixing parameters imply the occurrence of resonant oscillations near the neutron star which would swap $\nu_e \leftrightarrow \nu_\mu$ and would thus invert the energy hierarchy to $\langle E_{\nu_e} \rangle > \langle E_{\overline{\nu}_e} \rangle$, causing a proton-rich wind [13]. With inverted masses $m_{\nu_e} > m_{\nu_\mu}$ resonant oscillations occur among anti-neutrinos instead so that $\overline{\nu}_e \leftrightarrow \overline{\nu}_\mu$. The resulting energy hierarchy $\langle E_{\nu_e} \rangle < \langle E_{\overline{\nu}_\mu} \rangle < \langle E_{\overline{\nu}_e} \rangle = \langle E_{\nu_\mu} \rangle = \langle E_{\nu_\tau} \rangle = \langle E_{\overline{\nu}_\tau} \rangle$ should be compatible with r-process nucleosynthesis.

A 2.5 eV mass for $\nu_e$ could conflict with Majorana mass bounds from neutrinoless $\beta\beta$ decay. Currently, the best limit comes from the Heidelberg-Moscow Germanium experiment with an upper limit of $m_{\nu_e} \lesssim 0.7$ eV [14]. A conflict is avoided if the $\nu_e$ mass is of Dirac type, a solution which requires one to postulate the existence of low-mass right-handed neutrinos. More interestingly, one may retain Majorana masses if one recalls that what is really constrained in $\beta\beta$ decay experiments is the quantity

$$\langle m_{\nu_e} \rangle = \sum_{j=1}^{3} \lambda_j |U_{ej}|^2 m_j \quad (2)$$

where $m_j$ ($j = 1, 2, 3$) are the mass eigenstates, $U_{ej}$ their admixture to $\nu_e$, and $\lambda_j = \pm 1$ a CP-phase. By assumption, $m_2 \approx 0$ and $m_1 \approx m_3 \approx 2.5$ eV, so that a negative relative 1-3 CP phase allows for a cancellation if the 1-3 mixing angle is large. Thus, it is enough to require $\langle m_{\nu_e} \rangle = 2.5 \, \text{eV}(\cos^2 \theta - \sin^2 \theta) < 0.7$ eV where we used $U_{e1} = \cos \theta$ and $U_{e3} = \sin \theta$ with $\theta$ the 1-3 mixing angle which in our scenario is the one relevant for solar neutrino oscillations. Therefore, it is required that $\sin^2 2\theta > 0.92$, i.e., almost maximum mixing. However, the uncertainty of the nuclear matrix elements entering the Majorana mass bound may be as large as a factor of 2 [14] whence the limit may be as weak as $\langle m_{\nu_e} \rangle \lesssim 1.4$ eV so that only $\sin^2 2\theta \gtrsim 0.69$ would be needed.

The solar neutrino problem is solved by matter-induced oscillations if $\Delta m^2 \approx 10^{-5}$ eV$^2$ and either $\sin^2 2\theta \approx 0.007$ or $\sin^2 2\theta \approx 0.6$ (e.g. Hata and Langacker [15]), assuming a standard solar model with physical input parameters within their recognized experimental or systematic uncertainties. Therefore, our scenario favors the large-angle solution. It would cause a pronounced day-night effect in the Superkamiokande detector which hopefully will commence operation in the spring of 1996; within a few months of data taking this diurnal signal variation could be identified [16]. If this were the case one would be led to speculate that the discovery of neutrinoless $\beta\beta$ decay must be imminent.

Another large-angle solution is provided by vacuum oscillations with $\Delta m^2 \approx 10^{-10}\,\text{eV}^2$ (e.g. Krastev and Petcov [17]). In this case the oscillation length is of order the annual distance variation between Sun and Earth so that ultimately this solution can be identified by a semiannual variation of the measured fluxes. (Because the flux from the monochromatic beryllium line would show a particularly conspicuous time variation this effect could be observed well at BOREXINO which is the only forthcoming solar neutrino detector sensitive to beryllium neutrinos.) The vacuum solution is very attractive in the sense that it allows for maximum mixing so that the $\beta\beta$ decay limits will never cause a problem. In fact, one would view $\nu_e$ and $\nu_\tau$ as two Majorana components of one Dirac neutrino except for the small 1-3 mass difference which presumably would have to be explained by some higher-order correction to whichever physical effect that causes the neutrinos to have masses in the first place.

The atmospheric neutrino anomaly remains unexplained by oscillations in our scenario unless one invokes a sterile neutrino for this purpose. A far worse problem, however, is posed by the neutrino signal that was observed from SN 1987A in the IMB and Kamiokande II water Cherenkov detectors. Our inverted mass hierarchy avoids a "level crossing" between $\nu_e$ and $\nu_\mu$ on the way out of the supernovae, and so $\nu_e \leftrightarrow \nu_\mu$ exchange is avoided, a process that would have suppressed r-process nucleosynthesis. By the same token, however, a level crossing does occur between $\overline{\nu}_e$ and $\overline{\nu}_\mu$, and so their spectra are expected to be swapped. Therefore, the detectors should have measured higher-energy $\overline{\nu}_e$'s than have been observed.

A detailed maximum likelihood analysis of the SN 1987A neutrino signal has been performed by Loredo and Lamb [19]. Among ten different ways to parametrize the SN neutrino flux they favored a cooling model where neutrinos are emitted from a sphere at a fixed radius with a temperature which falls exponentially in time, beginning with an initial value $T_0$. For the present discussion the temporal evolution of the neutrino flux is of minor interest; we are mostly concerned with the fluence (time-integrated flux) at Earth as a function of the $\overline{\nu}_e$ energy. If one assumes that the instantaneous neutrino spectrum is given by a Fermi-Dirac distribution with a vanishing chemical potential, the exponential cooling model yields $\langle E_{\overline{\nu}_e} \rangle = 2.36\, T_0$ for the time-integrated spectrum. With the help of this conversion factor we show in Fig. 1 Loredo and Lamb's confidence contours for $\langle E_{\overline{\nu}_e} \rangle$ and the inferred total amount of emitted $\overline{\nu}_e$ energy. Typical SN calculations yield $E_{\text{tot}}(\overline{\nu}_e) \approx 0.5 \times 10^{53}$ erg, in good agreement with the best-fit value of Fig. 1. The inferred best-fit value of $\langle E_{\overline{\nu}_e} \rangle = 10.5\,\text{MeV}$ is somewhat low relative to the predictions cited in Eq. (1).

In Fig. 2 we show the expected spectra at the two detectors. The solid lines correspond to the exponential cooling model with the best-fit values $\langle E_{\overline{\nu}_e} \rangle = 10.5\,\text{MeV}$ ($T_0 = 4.5\,\text{MeV}$) and $E_{\text{tot}}(\overline{\nu}_e) = 0.5 \times 10^{53}$ erg. It agrees well with the dotted line which was derived from the numerical cooling calculation of Burrows [20], his model 55, for which the equation of state and matter accretion rate were adjusted such as to optimize agreement with the detector signals. In this model $\langle E_{\overline{\nu}_e} \rangle = 11.0\,\text{MeV}$; the instantaneous neutrino spectrum was taken to be black-body. This comparison confirms that Loredo and Lamb's parametrization is a reasonable representation of what one might expect from a numerical cooling calculation, at least for the time-integrated spectrum.

So far it was assumed that the instantaneous neutrino spectrum is a Fermi-Dirac distribution with vanishing chemical potential. However, the spectrum is probably "pinched" (suppressed low- and high-energy tails) as shown, e.g., by the Monte Carlo transport calculations of Janka and Hillebrandt [21]. These authors [22] have performed a maximum likelihood analysis of the SN 1987A signal on the basis of a spectral parametrization of the fluence in terms of an effective temperature and an effective degeneracy parameter $\eta$ where $\eta > 0$ corresponds to pinching (suppressed high energy tail), $\eta < 0$ to anti-pinching. For each detector separately, they found a best-fit value of $\eta = 0$; probably they would have found a negative value had they not imposed the restriction $\eta \geq 0$. For $\eta = 0$ their best-fit results are: $\langle E_{\overline{\nu}_e} \rangle = 9.1\,\text{MeV}$ and $E_{\text{tot}}(\overline{\nu}_e) = 0.64 \times 10^{53}$ erg (Kamiokande II) and $\langle E_{\overline{\nu}_e} \rangle = 14.2\,\text{MeV}$ and $E_{\text{tot}}(\overline{\nu}_e) = 0.34 \times 10^{53}$ erg (IMB). The confidence contours are not very restrictive on $\eta$. Fixing $\eta = 3.5$ on the basis of what is theoretically preferred, the best-fit results are: $\langle E_{\overline{\nu}_e} \rangle = 10.9\,\text{MeV}$ and $E_{\text{tot}}(\overline{\nu}_e) = 0.53 \times 10^{53}$ erg (Kamiokande II) and $\langle E_{\overline{\nu}_e} \rangle = 18.6$ MeV and $E_{\text{tot}}(\overline{\nu}_e) = 0.23 \times 10^{53}$ erg (IMB). Even for $\eta = 0$ the best-fit $\langle E_{\overline{\nu}_e} \rangle$ disagree substantially between Kamiokande II and IMB; their 95% confidence contours barely overlap. This discrepancy is worse for the assumed pinching with $\eta = 3.5$. The data prefer anti-pinching in each detector separately, and to explain the relatively large number of events at IMB (which is only sensitive to the high-energy tail) relative to Kamiokande II. Of course, anti-pinching of the time-integrated spectrum is not necessarily in conflict with the assumption of pinched instantaneous fluxes.

The spectral distribution at the detectors lead to values for $\langle E_{\overline{\nu}_e} \rangle$ which are slightly low, but compatible with predictions of numerical calculations. However, at a reasonable confidence level they are not compatible with the assumption that primarily converted $\overline{\nu}_\mu$'s had been observed. For $\overline{\nu}_\mu$'s the expected spectral pinching is far less pronounced than for $\overline{\nu}_e$'s so that the Loredo and Lamb analysis is essentially adequate. Their joint analysis between both detectors allows $\langle E_{\overline{\nu}_e} \rangle = 15.5$ MeV ($T_0 = 6.5$ MeV) with $E_{\text{tot}}(\overline{\nu}_e) = 0.2 \times 10^{53}$ erg just barely at the 95% CL. From Fig. 2 (dashed line) the disagreement with the Kamiokande II signal is plainly visible. Even this large value for $\langle E_{\overline{\nu}_e} \rangle$ is far below the expectation for converted $\overline{\nu}_\mu$'s. Therefore, the SN 1987A signal favors the normal over the inverted mass hierarchy by a large margin.

The predictions of the spectral features of the neutrino signal are fraught with uncertainties. One is related to the neutrino opacities at high density where one expects that fast spin-fluctuations lead to a suppression of the axial-vector scattering rate at densities beyond, say, 10% nuclear [23]. The magnitude of this effect has not yet been calculated from first principles. Instead, the neutrino opacities have recently been "calibrated" from the SN 1987A signal [24]. Reduced neutrino opacities lead to a stiffening of the observable $\overline{\nu}_e$ spectrum [24] and so, the average energies of Eq. (1) which were predicted on the basis of naive unsuppressed scattering cross sections likely are lower limits. Therefore, it is even less believable that converted $\overline{\nu}_\mu$'s were observed from SN 1987A.

Does one really expect a complete swap $\overline{\nu}_e \leftrightarrow \overline{\nu}_\mu$ by matter-induced oscillations? For our inverted mass scheme the $\overline{\nu}_e$ becomes degenerate with the $\overline{\nu}_\mu$ when $\Delta m^2 / 2 E_\nu = \sqrt{2} G_{\text{F}} n_e$ with the Fermi constant $G_{\text{F}}$ and the electron density $n_e$. With the LSND value $\Delta m^2 = 6 \, \text{eV}^2$, with $E_\nu = 30$ MeV which is characteristic of the detected neutrinos, and with an electron frac-tion of 0.5 per baryon one finds that the cross-over occurs at a density of $2.6 \times 10^6\,\mathrm{g\,cm^{-3}}$. This is far outside the neutrino sphere which is at a density of around $10^{13}\,\mathrm{g\,cm^{-3}}$. The oscillations are adiabatic if on resonance $|\nabla \ln n_e| \gg (\Delta m^2/2E_\nu)\sin 2\theta\,\tan 2\theta$. With the LSND value $\sin^2 2\theta = 0.005$ and $E_\nu = 30\,\mathrm{MeV}$, this condition reads $|\nabla \ln n_e|^{-1} \gg 0.4\,\mathrm{km}$, which is easily satisfied even after the shock wave has expelled the stellar mantle so that the density falls off relatively rapidly near the neutron star surface.

Of course, at relatively late times the resonance occurs at a few 10 km from the neutron star surface while at early times, before the shock wave has escaped, it occurs at a few 1000 km. Clearly, there will be complicated dynamics and perhaps large-scale motions of material and turbulence which render a simple spherically symmetric density profile to be too simplistic. While it appears implausible that the adiabatic condition will be destroyed for a substantial region of space or a significant period of time during the Kelvin-Helmholtz cooling phase of the neutron star, a more detailed analysis may well be called for before passing final judgement.

Besides the electron density profile, one has to worry about neutrino self-interactions with regard to the matter-induced oscillation problem. The same issue arises for the normal mass hierarchy and the resulting $\nu_e \leftrightarrow \nu_\mu$ oscillations, a problem that was studied in Ref. [13] in the context of a possible suppression of the r-process nucleosynthesis. The neutrino self-interactions seem to cause only a mild modification of the adiabatic conversion process [13]. Still, these issues have only begun to be fully understood, and so one may not be entirely satisfied yet with the approximate treatments that are hitherto available.

The interpretation that the SN 1987A signal was indeed caused by $\overline{\nu}_e$'s by virtue of the $\overline{\nu}_e p \to n e^+$ reaction is in itself subject to doubt because instead of the expected isotropic signature one found a forward peaked signal, especially for the high-energy events, which is compatible with isotropy at far less than a 1% probability [25]. Because there is no plausible alternative explanation available at the present time one must settle for the interpretation of a rare statistical fluctuation of the angular distribution of the events, a fluctuation which should not affect the interpretation of the average energies and number of events. Still, one remains uneasy about far-reaching conclusions which rely on a signal with rather extreme anomalous features.

In spite of these systematic uncertainties associated with the interpretation of the SN 1987A neutrino signal it is difficult to simply discard it in favor of our inverted mass scenario. Is there another way out? One possibility is indicated by the observation that the LSND experiment alone does not allow one to fix $\Delta m^2$ with certainty. The $\ell/E$ distribution of the events which are associated with the $\overline{\nu}_\mu \to \overline{\nu}_e$ channel also allow for the values 0.5, 2, 6, or 10 eV$^2$ with a reasonable significance [2]. The value 6 eV$^2$ was favored because of a sensitivity dip of KARMEN and BNL-E776 [2, 3]. However, if the LSND-favored $\sin^2 2\theta$ is indeed below 0.01, contrary to early discussions, there may be no conflict with KARMEN at any $\Delta m^2$ [6]. This leaves a potential compatibility problem with BNL-E776 for which two different exclusion plots are in circulation. One[1] shows a sensitivity minimum at $\Delta m^2 = 6$ eV$^2$; it requires $\sin^2 2\theta < 0.005$ at 90% CL. In contrast, the exclusion plot in the official BNL-E776 publication [5] shows a sensitivity minimum at 8 eV$^2$ (a value disfavored by LSND) where $\sin^2 2\theta < 0.003$ is required at 90% CL. It shows high sensitivity at 2, 6, and 10 eV$^2$ where $\sin^2 2\theta \lesssim 0.002$. This nominally excludes most of the LSND-favored values unless the errors of BNL-E776 were larger than stated, or if one appeals to a rare statistical fluctuation. On the other hand, if the LSND results were explained by $\Delta m^2 \approx 0.5$ eV$^2$ there does not seem to be even a nominal disagreement with KARMEN or BNL-E776 even for mixing angles as large as $\sin^2 2\theta \approx 0.02$. It remains to be seen if the putative evidence for oscillations in the $\nu_\mu \to \nu_e$ channel at LSND selects any of the $\Delta m^2$ values with superior significance.

On the face of it, a low $\Delta m^2$ destroys the beautiful cosmological scenario of Primack et al. [3]. However, it is only required that the sum of the neutrino masses is about 5 eV. Together with $|m_{\nu_e}^2 - m_{\nu_\mu}^2| \approx 0.5$ eV$^2$ one could contemplate a mass matrix $m_{\nu_e} \approx m_{\nu_\tau} \approx 1.62$ eV and $m_{\nu_\mu} \approx 1.77$ eV. This C$\nu^3$DM model has an interesting implication for cosmology: it provides a dip in the ratio of the power spectrum relative to that for the 1 neutrino model, at fixed $\Omega_\nu$, at approximately twice the scale found in a similar comparison for the 2-neutrino model, that is, near $20h^{-1}$ Mpc, by analogy with the computations cited in reference [3]. This would help further suppress the abundance of massive galaxy clusters and amplitudes of large-scale flows, and thereby lead to better consistency with the observational constraints [27].

---

[1] This plot has appeared in the Proceedings of Neutrino 90 (pg. 148), in the review [26], in the KARMEN paper [6], and was recently shown at the Moriond conference in Villars, Switzerland, January 1995. We understand that this plot was based on a narrow-band beam, while the 1992 publication on a wide-band beam (W. Lee, communicated to us by D. Caldwell). Therefore, the 1992 version [5] should be used.

Even with the proposed normal $\nu_e$-$\nu_\mu$-hierarchy the r-process nucleosynthesis is left unscathed as the resonant oscillations would now occur too far away from the neutron star surface; even the value $\Delta m^2 \approx 6\,\mathrm{eV}^2$ was close to the edge of the sensitive regime [13].

An explanation of the solar neutrino problem by $\nu_e \to \nu_\tau$ oscillations still requires a large mixing angle in order to avoid a conflict with the $\beta\beta$ Majorana mass limits. However, with $m_{\nu_e} \approx m_{\nu_\tau} \approx 1.62\,\mathrm{eV}$ the nominal Germanium bound $\langle m_{\nu_e}\rangle < 0.7\,\mathrm{eV}$ translates into $\sin^2 2\theta > 0.8$ so that both large-angle solutions are possible without the need to appeal to large uncertainties of the nuclear matrix elements.

With $m_{\nu_e} < m_{\nu_\mu}$ there is no problem with the SN 1987A signal. On the contrary, if the solar neutrino problem is solved by the large-angle MSW solution the $\nu_e$-$\nu_\tau$ mixing parameters are such that matter-induced oscillations in the Earth are important; this effect leads to the diurnal signal variation that would be expected at Superkamiokande. For SN 1987A, the detectors would have observed a certain combination of primary $\overline{\nu}_e$'s with $\overline{\nu}_\tau$'s whose spectra would have been partially swapped by the Earth effect. As Kamiokande II looked at SN 1987A through approx. 3900 km of the Earth, IMB through approx. 8400 km, the effective spectrum observed by IMB would have been stiffened relative to Kamiokande II, partially explaining the "tension" between the best-fit $\langle E_{\overline{\nu}_e}\rangle$ inferred from the two detectors [28]. In our first mass scheme with $m_{\nu_e} > m_{\nu_\mu}$ this effect would not change the relative spectra because the $\overline{\nu}_e$'s arriving at Earth would have been converted $\overline{\nu}_\mu$'s with a spectrum identical to that of the $\overline{\nu}_\tau$'s.

In summary, if the LSND neutrino survives further experimental scrutiny, an explanation of the solar neutrino problem by oscillations among sequential neutrinos requires a near $\nu_e$-$\nu_\tau$ degeneracy, and so requires either a mass inversion between $\nu_e$ and $\nu_\mu$ or between $\nu_\mu$ and $\nu_\tau$. If the LSND signature is explained by $\Delta m^2 = 6\,\mathrm{eV}^2$ as originally proposed, the r-process nucleosynthesis argument as well as the $C\nu^2\mathrm{DM}$ cosmology favor $m_{\nu_e} \approx m_{\nu_\mu} \approx 2.5\,\mathrm{eV}$ and $m_{\nu_\mu} \approx 0$, a scenario that is problematic in view of the SN 1987A neutrino observations. If the LSND value is $\Delta m^2 = 0.5\,\mathrm{eV}^2$, cosmology favors $m_{\nu_e} \approx m_{\nu_\tau} \approx 1.62\,\mathrm{eV}$ and $m_{\nu_\mu} \approx 1.77\,\mathrm{eV}$ which avoids any conflict with SN physics. Unless one appeals to neutrino Dirac masses, the $\beta\beta$ Majorana mass limits require in either scenario a large-angle solution to the solar neutrino problem which likely can be identified at the forthcoming solar neutrino experiments by a diurnal or semiannual time variation.

**Acknowledgements:** G.R. enjoyed the hospitality of the N.S.F. Center for Particle Astrophysics in Berkeley in the fall of 1994 when this work was begun. Our research at Berkeley is also supported in part by the U.S. Department of Energy. We thank D. O. Caldwell (LSND collaboration), F. von Feilitzsch (BOREXINO collaboration), J. Kleinfeller (KARMEN collaboration), H. V. Klapdor-Kleingrothaus (Moscow-Heidelberg $\beta\beta$ experiment), and H. Meyer (Fréjus Collaboration) for discussions and information on their experiments. We have benefitted from remarks by G. Sigl, W. Hillebrandt and G. Fuller on early drafts of this paper. We are particularly indebted to H.-Th. Janka for a lengthy electronic correspondence on the matter of expected SN neutrino spectra. We are aware that G. Fuller, J. Primack and Y.-Z. Qian have also investigated the consequences of an inverted mass scheme.

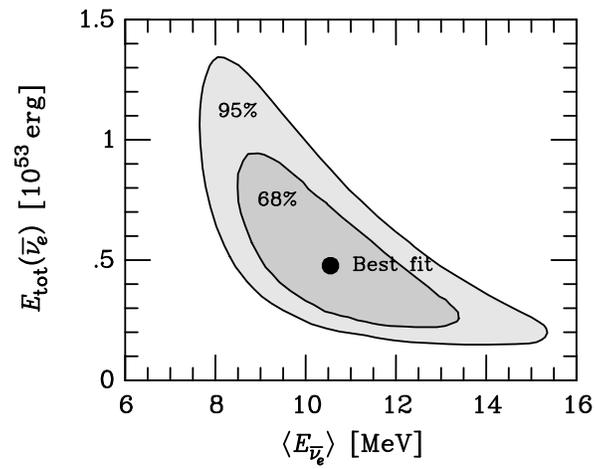

Figure 1: Projection of the confidence volume in a maximum likelihood analysis of the SN 1987A neutrino signal, assuming an exponential cooling model (adapted from Loredo and Lamb [19] with our conversion factor $\langle E_{\bar{\nu}_e}\rangle = 2.36\, T_0$).

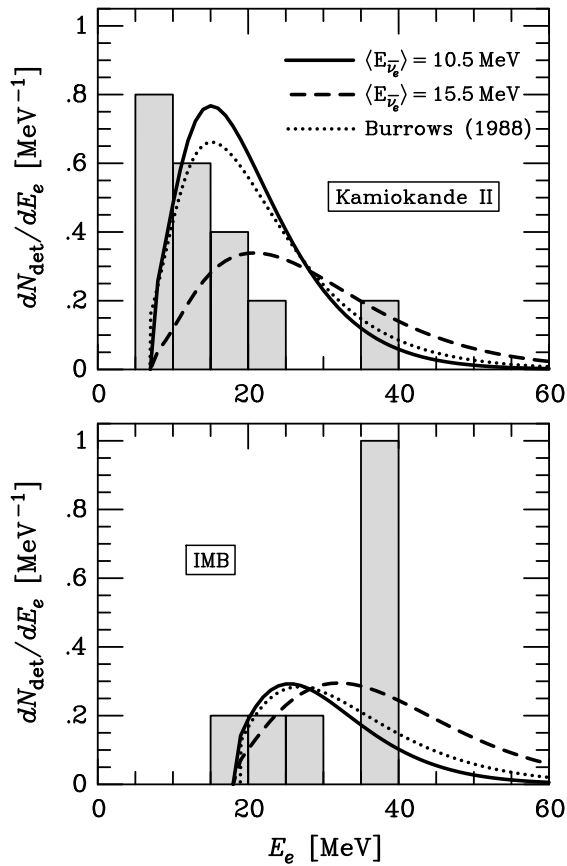

Figure 2: Expected and measured distributions of the events in the Kamiokande II and IMB detectors from SN 1987A. On the vertical axis, 1 event in the binned spectra corresponds to 0.2. The solid and dashed lines are predictions for exponential cooling models. For $\langle E_{\bar{\nu}_e}\rangle = 10.5\,\mathrm{MeV}$, which corresponds to the best-fit value of Fig. 1, a total energy of $0.5\times 10^{53}$ erg was assumed to have been emitted in $\bar{\nu}_e$'s. For $\langle E_{\bar{\nu}_e}\rangle = 15.5\,\mathrm{MeV}$ we chose $E_{\mathrm{tot}}(\bar{\nu}_e) = 0.2\times 10^{53}$ erg, corresponding to the right-hand tip of the 95% contour in Fig. 1. The dotted line is the prediction from the model 55 calculation of Burrows [20].

14